\documentclass{ws-procs11x85}

\makeindex
\begin{document}

\title{QCD AND HEAVY HADRON DECAYS\footnote{}}

\author{G. BUCHALLA}

\address{Ludwig-Maximilians-Universit\"at M\"unchen, Sektion Physik,
Theresienstra\ss e 37, D-80333 M\"unchen, Germany
\\E-mail: Gerhard.Buchalla@physik.uni-muenchen.de}


\twocolumn[\maketitle\abstract{
We review recent developments in QCD pertaining to its
application to weak decays of heavy hadrons.
We concentrate on exclusive rare and nonleptonic $B$-meson decays,
discussing both the theoretical framework and phenomenological
issues of current interest.}]

\baselineskip=13.07pt

\footnotetext{$^*$Talk at Lepton Photon 03, 11-16 August 2003, 
Fermilab, Batavia, USA; preprint LMU 01/04}

\section{Introduction}
Weak decays of heavy hadrons, of $B$ mesons in particular, provide
us with essential information on the quark flavor sector. Since the
underlying flavor dynamics of the quarks is masked by strong
interactions, a sufficiently precise understanding of QCD effects
is crucial to extract from weak decays involving hadrons the basic
parameters of flavor physics.
Much interest in this respect is being devoted to rare $B$ decay modes
such as $B\to\pi\pi$, $\pi K$, $\pi\rho$, $\phi K_S$, $K^*\gamma$,
$\rho\gamma$ or $K^* l^+l^-$. These decays are a rich source of information
on CKM parameters and flavor-changing neutral currents. Many new results
are now being obtained from the $B$ meson factories and hadron colliders.
\cite{TB,JF,HJ,MN,KP,KS}
Both exclusive and inclusive decays can be studied.
Roughly speaking, the former are more difficult for theory, the
latter for experiment. 

In dealing with the presence of strong
interactions in these processes the challenge for theory is
in general to achieve a systematic separation of long-distance
and short-distance contributions in QCD. This separation typically
takes the form of representing an amplitude or a cross section
as a sum of {\it products} of long and short distance quantities
and is commonly refered to as {\it factorization\/}. 
The concept of factorization requires the existence of at least
one hard scale, which is large in comparison with the intrinsic scale
of QCD. For $B$ decays this scale is given by the $b$-quark mass,
$m_b\gg\Lambda_{QCD}$.
The asymptotic freedom of QCD allows one to compute the short-distance
parts using perturbation theory. Even though the long-distance
quantities still need to be dealt with by other means, the procedure
usually entails a substantial simplification of the problem.

Various methods, according to the specific nature of
the application, have been developed to implement the idea of
factorization in the theoretical description of heavy hadron
decays. These include heavy-quark effective theory (HQET),
heavy-quark expansion (HQE), factorization in exclusive nonleptonic
decays and soft-collinear effective theory (SCET). In particular the latter
two topics are more recent developments and are still under active
investigation and further study. They play an important role for
the exclusive rare $B$ decays listed above.
Dynamical calculations based on these tools hold the promise to improve our
understanding of QCD in heavy-hadron decays significantly and to
facilitate the determination of fundamental weak interaction
parameters. A different line of approach is the use of the
approximate $SU(2)$ or $SU(3)$ flavor symmetries of QCD in order to
isolate the weak couplings in a model-independent way.\cite{GL,GHLR,FD,CGLRS} 
Both strategies, flavor symmetries and dynamical calculations, are
complementary to each other and enhance our ability to test
quark flavor physics. While the flavor symmetry approach gives constraints
free of hadronic input in the symmetry limit, dynamical methods allow us
to compute corrections from flavor symmetry breaking.

The following section gives a brief overview of theoretical frameworks
for $B$ decays based on the heavy-quark limit. The remainder of this
talk then concentrates on the subject of exclusive rare or
hadronic decays of $B$ mesons.

\section{Tools and Applications}

\begin{figure}
[htbp]
\centering
\includegraphics[width=2.5cm,height=1.5cm]{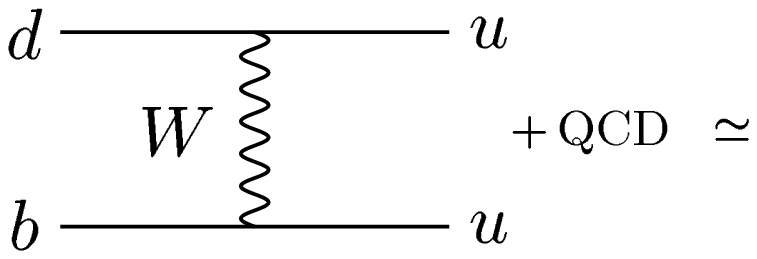}
\raisebox{0.7cm}{{$C\left(\frac{M_W}{\mu}, \alpha_s\right)\cdot$}}
\includegraphics[width=2cm,height=1.5cm]{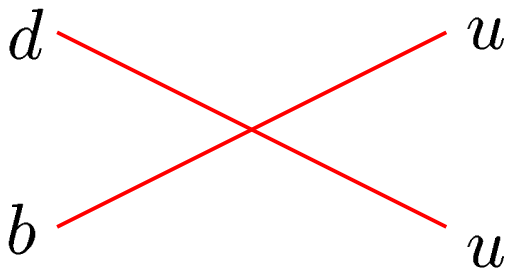}
\caption[]{OPE for weak decays.}
\label{fig:ope}
\end{figure}

The application of perturbative QCD to hadronic reactions at high
energy requires a proper factorization
of short-distance and long-distance contributions.
One example is given by the operator product expansion (OPE)
used to construct effective Hamiltonians for hadronic $B$ decay.
This is shown schematically in Fig. \ref{fig:ope} for a generic
$B$ decay amplitude. 
The OPE approximates the nonlocal product of two weak currents,
which are connected by $W$ exchange in the full standard model,
by local 4-quark operators, multiplied by Wilson coefficients
$C(M_W/\mu,\alpha_s)$. In this way the short-distance physics
from scales of order $M_W$ (or $m_t$ appearing in penguin loop diagrams)
down to a factorization scale $\mu\sim m_b$ is isolated into the
coefficient. Determined by high enrgy scales, the coefficient can
be computed perturbatively, supplemented by renormalization-group
improvement to resum large logarithms $\sim\alpha_s\ln M_W/m_b$.
The QCD dynamics from scales below $\mu$ is contained within the
matrix elements of the local operators. These matrix elements depend
on the particular process under consideration, whereas the coefficients
are universal. The approximation is valid up to power corrections of 
order $m^2_b/M^2_W$.

In the case of $B$ decay amplitudes, the hadronic matrix elements
themselves still contain a hard scale $m_b\gg\Lambda_{QCD}$.
Contributions of order $m_b$ can be further factorized from the
intrinsic long-distance dynamics of QCD. This is implemented by
a systematic expansion in $\Lambda_{QCD}/m_b$ and $\alpha_s(m_b)$
and leads to important simplifications.
The detailed formulation of this class of factorization depends on
the specific application and can take the form of
HQET, HQE, QCD factorization for exclusive hadronic $B$ decays
or SCET.

\begin{itemize}
\item
HQET describes the static approximation for a heavy quark, 
formulated in a covariant way as an effective field theory.\cite{IW,MN93}
It allows for a systematic inclusion of power corrections.
Its usefulness is based on two important features: The spin-flavor
symmetry of HQET relates form factors in the heavy-quark limit
and thus reduces the number of unknown hadronic quantities. Second, 
the dependence on the heavy-quark mass is made explicit.
Typical applications are (semi)leptonic form factors involving
hadrons containing a single heavy quark, such as $B\to D^{(*)}$
form factors in semileptonic $b\to c$ transitions or the decay
constant $f_B$.
\item
HQE is a theory for inclusive $B$ decays.\cite{BBSUV,BSU} 
It is based on the optical
theorem for inclusive decays and an operator product expansion
in $\Lambda_{QCD}/m_b$ of the transition operator. The heavy-quark
expansion justifies the parton model for inclusive decays of heavy hadrons,
which it contains as its first approximation. Beyond that it allows
us to study nonperturbative power corrections to the partonic picture.
The main applications of the HQE method is for processes as
$B\to X_{u,c}l\nu$, $B\to X_s\gamma$, $B\to X_s l^+l^-$, and
for the lifetimes of $b$-flavored hadrons.
\item
QCD factorization refers to a framework for analysing exclusive
hadronic $B$ decays with a fast light meson as for instance
$B\to D\pi$, $B\to\pi\pi$, $B\to\pi K$ and $B\to V\gamma$. 
This approach is conceptually similar to the theory of
hard exclusive reactions, described for instance by the pion electromagnetic
form factor at large momentum transfer.\cite{ER,LB} 
The application to $B$ decays
requires new elements due to the presence of heavy-light mesons.\cite{BBNS}
\item
SCET is an effective field theory formulation for transitions 
of a heavy quark into an energetic light quark.\cite{BFPS} The basic idea
is reminiscent of HQET. However, the structure of SCET is more
complex because the relevant long-distance physics
that needs to be factorized includes both soft and collinear
degrees of freedom. Only soft contributions have to be accounted
for in HQET. 
Important applications of SCET are the study of $B\to P$, $V$
transition form factors at large recoil energy of the light
pseudoscalar ($P$) or vector ($V$) meson, and formal proofs
of QCD factorization in exclusive heavy hadron decays.
\end{itemize}

There are further methods, which have been
useful to obtain information on hadronic quantities relevant to
$B$ decays. Of basic importance are computations based on
lattice QCD, which can access many quantities needed for
$B$ meson phenomenology (see \cite{ASK} for a recent review).
On the other hand, exclusive processes with fast light particles
are very difficult to treat within this framework. An important
tool to calculate in particular heavy-to-light form factors ($B\to\pi$)
at large recoil are QCD sum rules on the light cone.\cite{PBB,BKR}
We will not discuss those methods here, but refer to the 
literature for more information.

\section{Exclusive Hadronic $B$ Decays in QCD}

\subsection{Factorization}

The calculation of $B$-decay amplitudes, such as 
$B\to D\pi$, $B\to\pi\pi$ or $B\to \pi K$, starts from
an effective Hamiltonian, which has, schematically, the form
\begin{equation}
{\cal H}_{eff}=\frac{G_F}{\sqrt{2}}\lambda_{CKM}\, C_i Q_i
\end{equation}
Here $C_i$ are the Wilson coefficients at a scale $\mu\sim m_b$,
which are known at next-to-leading order in QCD.\cite{BBL}
$Q_i$ are local, dimension-6 operators and $\lambda_{CKM}$ represents
the appropriate CKM matrix elements. The main theoretical problem
is to evaluate the matrix elements of the operators $\langle Q_i\rangle$
between the initial and final hadronic states. A typical matrix element
reads $\langle\pi\pi|(\bar ub)_{V-A}(\bar du)_{V-A}|B\rangle$.

\begin{figure*}[t]
\center
\vspace{-3cm}
\psfig{figure=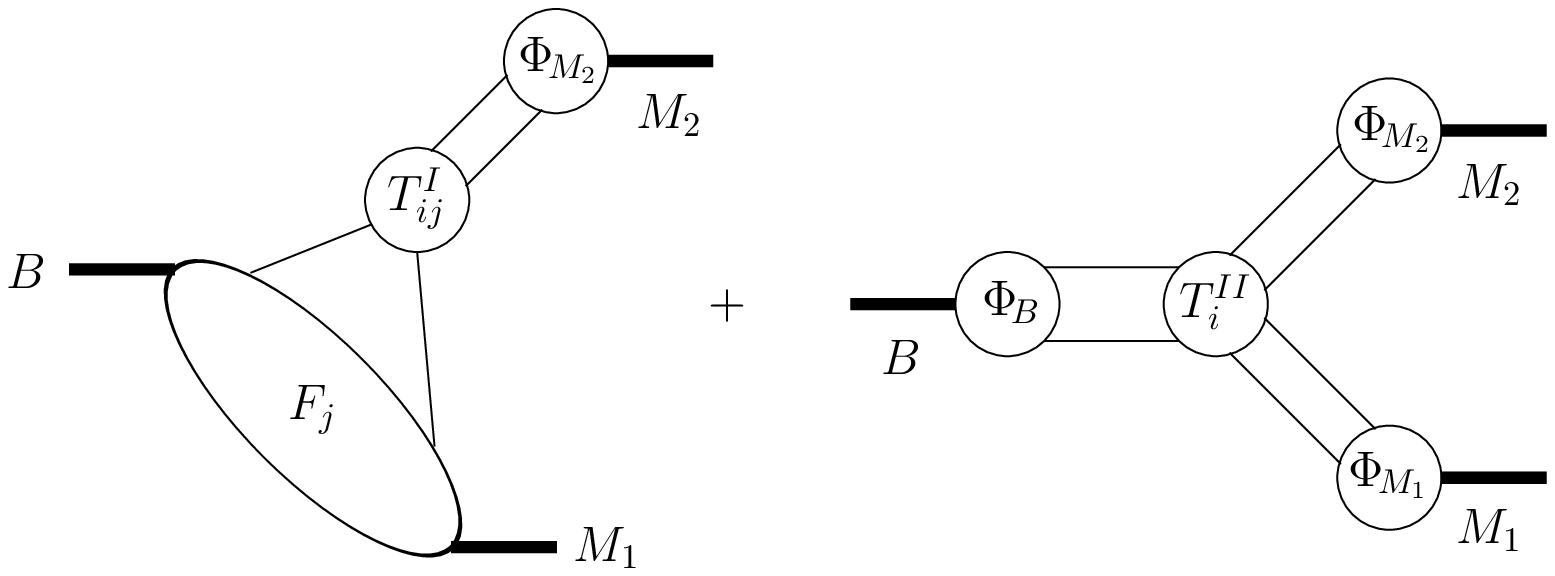,width=6.6truein,height=8.0truein}
\vspace*{-13cm}
\caption[]{Graphical representation of the factorization
formula. \label{fig:fform}}
\end{figure*}

These matrix elements simplify in the heavy-quark limit, where
they can in general be written as the sum of two terms, each
of which is factorized into hard scattering functions $T^I$ and $T^{II}$,
respectively, and the nonperturbative, but simpler, form factors
$F_j$ and meson light-cone distribution amplitudes $\Phi_M$ 
(Fig. \ref{fig:fform}).

Important elements of this approach are: i) The expansion
in $\Lambda_{QCD}/m_b\ll 1$, consistent power counting, and
the identification of the leading power contribution, for which the
factorized picture can be expected to hold.
ii) Light-cone dynamics, which determines for instance the
properties of the fast light mesons. The latter are described by
light-cone distribution amplitudes $\Phi_\pi$ of their
valence quarks defined as 
\begin{equation}
\langle\pi(p)|u(0)\bar d(z)|0\rangle=\frac{if_\pi}{4}\ \gamma_5\not\! p
\ \int^1_0 dx\ e^{ix pz}\ \Phi_\pi(x)
\end{equation}
with $z$ on the light cone, $z^2=0$.
iii) The collinear quark-antiquark pair dominating the
interactions of the highly energetic pion decouples from soft
gluons (colour transparency). This is the intuitive reason behind
factorization. iv) The factorized amplitude consists of hard, short- 
distance components, and soft, as well as collinear, long-distance
contributions.
More details on the factorization formalism can be found
elsewhere \cite{BBNS}.

An alternative approach to exclusive two-body decays
of $B$ mesons, refered to as pQCD, has been proposed in \cite{KLS}.
The main hypothesis in this method is that the $B\to\pi$ form
factor is not dominated by soft physics, but by hard gluon
exchange that can be computed perturbatively. The hypothesis
rests on the idea that Sudakov effects will suppress soft
endpoint divergences in the convolution integrals.
A critical discussion of this framework has been given in \cite{DGS2}.

\subsection{CP Violation in $B\to\pi^+\pi^-$}

\begin{figure}
\center
\psfig{figure=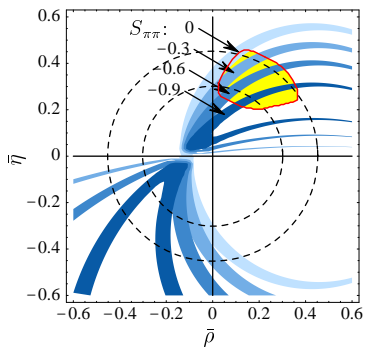, width=10truecm}
\caption[]{Constraints in the $\bar\rho$, $\bar\eta$ plane
from CP violation observable $S$ in $B\to\pi^+\pi^-$.
The constraints from $|V_{ub}/V_{cb}|$ (dashed circles)
and from the standard analysis of the unitarity triangle
(irregular shaded area) are also shown.} 
\label{fig:spipi}
\end{figure}

A framework for systematic computations of heavy-hadron
decay amplitudes in a well-defined limit clearly has
many applications for quark flavor physics with two-body
nonleptonic $B$ decays. 
An important example may serve to illustrate this point.
Consider the time-dependent, mixing-induced 
CP asymmetry in $B\to\pi^+\pi^-$
\begin{eqnarray}
{\cal A}_{CP}(t) &=&
\frac{\Gamma(B(t)\to\pi^+\pi^-)-\Gamma(\bar B(t)\to\pi^+\pi^-)}{
      \Gamma(B(t)\to\pi^+\pi^-)+\Gamma(\bar B(t)\to\pi^+\pi^-)}\nonumber  \\
&=& -S \sin(\Delta M_d t)+ C \cos(\Delta M_d t)
\end{eqnarray}
Using CKM-matrix unitarity, the decay amplitude consists of
two components with different CKM factors and different hadronic
parts, schematically
\begin{eqnarray}
&& A(B\to\pi^+\pi^-) = \\
&& \quad V^*_{ub} V_{ud} ({\rm up} - {\rm top})+
  V^*_{cb} V_{cd} ({\rm charm} - {\rm top}) \nonumber
\end{eqnarray}
If the penguin contribution $\sim V^*_{cb} V_{cd}$ could be
neglected, one would have $C=0$ and $S=\sin 2\alpha$, hence a direct
relation of ${\cal A}_{CP}$ to the CKM angle $\alpha$.
In reality the penguin contribution is not negligible compared 
to the dominant tree contribution $\sim V^*_{ub} V_{ud}$.
The ratio of penguin and tree amplitude, which enters the
CP asymmetry, depends on hadronic physics.
This complicates the relation of observables $S$ and $C$
to CKM parameters. QCD factorization of $B$-decay matrix elements
allows us to compute the required hadronic input and to determine
the constraint in the ($\bar\rho$, $\bar\eta$) plane implied by
measurements of the CP asymmetry. This is illustrated for $S$
in Fig. \ref{fig:spipi}. 
The widths of the bands indicate the 
theoretical uncertainty \cite{BBNS3}. Note that the constraints
from $S$ are relatively insensitive to theoretical or
experimental uncertainties. The analysis of direct CP violation
measured by $C$ is more complicated due to the importance of
strong phases.
Recent phenomenological analyses were performed in \cite{GR02,BS}. 
The current experimental results for $S$ and $C$ are 
from BaBar \cite{BABAR0}
\begin{eqnarray}
S &=& +0.02\pm 0.34\pm 0.05 \\
C &=& -0.30\pm 0.25\pm 0.04 
\end{eqnarray}
and from Belle \cite{BELLE1}
\begin{eqnarray}
S &=& -1.23\pm 0.41 ^{+0.08}_{-0.07} \\ 
C &=& -0.77\pm 0.27\pm 0.08 
\end{eqnarray}
A recent preliminary update from BaBar gives \cite{HJ,HFAG}
\begin{eqnarray}\label{scprel}
S &=& -0.40\pm 0.22\pm 0.03 \\
C &=& -0.19\pm 0.19\pm 0.05
\end{eqnarray}
Including the new BaBar results the current world average
reads \cite{HJ}
\begin{equation}\label{scwa}
S=-0.58\pm 0.20 \qquad C=-0.38\pm 0.16
\end{equation}
which ignores the large $\chi^2$ reflecting the 
relatively poor agreement between the experiments.

\subsection{Current Status}

QCD factorization to leading power in $\Lambda/m_b$ has been
demonstrated at ${\cal O}(\alpha_s)$ for the important class
of decays $B\to\pi\pi$, $\pi K$. For $B\to D\pi$ (class I),
where hard spectator interactions are absent, a proof
has been given explicitly at two loops \cite{BBNS} and
to all orders in the framework of soft-collinear effective theory
(SCET) \cite{BPS}. Complete matrix elements are available
at ${\cal O}(\alpha_s)$ (NLO) for $B\to\pi\pi$, $\pi K$,
including electroweak penguins.\cite{BBNS3}
Comprehensive treatments have also been given for
$B\to PV$ modes \cite{BN} (see also \cite{AGMPS}) and
for $B$ decays into light flavor-singlet mesons \cite{BN02}.  
A discussion of two-body $B$ decays into light mesons within SCET
has been presented in \cite{CK}.

Power corrections are presently not calculable in general.
Their impact has to be estimated and included into the error
analysis. Critical issues here are annihilation
contributions and certain corrections
proportional to $m^2_\pi/((m_u+m_d)m_b)$, which is numerically
sizable, even if it is power suppressed.
However, the large variety of channels available will provide us with
important cross checks and arguments based on SU(2) or SU(3)
flavor symmetries can also be of use in further controling
uncertainties.

\subsection{Phenomenology of $B\to PP$, $PV$}

Two-body $B$ decays into light mesons have been widely discussed
in the literature.\cite{SWGR}

In general, a phenomenological analysis of these modes
faces the problem of disentangling three very different aspects, which 
simultaneously affect the observable decay rates and asymmetries:
First, there are the CKM couplings that one would like to extract
in order to test the standard model. Second, it is possible that
some observables could be significantly modified by new physics
contributions, which would complicate the determination of CKM phases.
Third, the short distance physics, CKM quantities
and potential new interactions, that one is aiming for,
is dressed by the effects of QCD.
A priori any discrepancy between data and expectations has to be
examined with these points in mind.
Fortunately, the large number of different channels with
different QCD dynamics and CKM dependence will be very helpful
to clarify the phenomenological interpretation.
The following examples illustrate how various aspects of the
QCD dynamics may be tested independently.

\begin{enumerate}
\item Penguin-to-tree ratio.
To test predictions of this ratio a useful observable can be
built from the mode $B^-\to\pi^- \bar K^0$, which is entirely
dominated by a penguin contribution, and from the pure tree-type
process $B^-\to\pi^-\pi^0$:
\begin{equation}\label{rtilde}
\left|\frac{penguin}{tree}\right|=
\left|\frac{V_{ub}}{V_{cb}}\right|\frac{f_\pi}{f_K}
\sqrt{\frac{B(B^-\to\pi^- \bar K^0)}{2\, B(B^-\to\pi^-\pi^0)}}
\end{equation}
This amplitude ratio is not identical to the $P/T$ ratio
required for $B\to\pi^+\pi^-$, but still rather similar to
be interesting as a test. Small differences come from $SU(3)$
breaking effects (the dominant ones due to $f_\pi/f_K$ are already
corrected for in (\ref{rtilde})), and weak annihilation corrections
in $B\to \pi K$, and from the color-suppressed contribution to
$B^-\to\pi^-\pi^0$. Because the $\pi^-\bar K^0$ and $\pi^-\pi^0$
channels have only a single amplitude (penguin or tree), no
interference is possible and the ratio in (\ref{rtilde}) is
independent of the CKM phase $\gamma$. This is useful for
distinguishing QCD effects from CKM issues.
A comparison of factorization predictions
for the left-hand side of (\ref{rtilde}) with data used to compute
the right-hand side in (\ref{rtilde}) is shown in Fig. \ref{fig:pt}.
The agreement is satisfactory within uncertainties.
\begin{figure}
\center
\psfig{figure=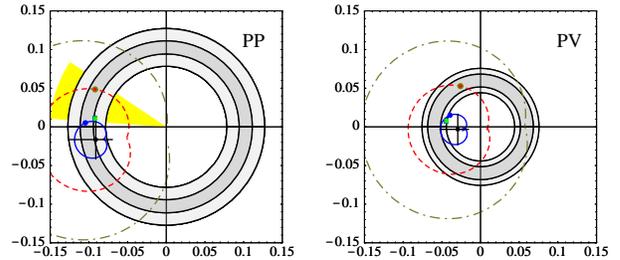, width=8truecm}
\caption[]{Left panel: Penguin-to-tree ratio extracted from data on
$B^-\to\pi^-\bar K^0$ and $B^-\to\pi^-\pi^0$ (rings in the center)
compared with predictions in QCD factorization (cross). The light
(dark) ring is with (without) the uncertainty from $|V_{ub}/V_{cb}|$.
The prediction includes a model estimate of power corrections, dominantly
from weak annihilation. The solid, dashed, dashed-dotted error contour
indicates the uncertainty from assigning $100\%$, $200\%$, $300\%$
error, respectively, to the default annihilation correction.
Right panel: The same with the $K$ replaced by $K^*$. (From \cite{BN})}
\label{fig:pt}
\end{figure}

\item Factorization test for $B^-\to\pi^-\pi^0$.
It is of interest to test predictions for the tree-amplitude
alone using a classical factorization test of the form
\begin{eqnarray}\label{pp0sl}
&& B(B^+\to\pi^+\pi^0) = 3\pi^2 f^2_\pi |V_{ud}|^2 \times\\
&& \quad \left. \frac{dB(B_d\to\pi^- l^+\nu)}{dq^2}\right|_{q^2=0}\,
\frac{\tau(B^+)}{\tau(B_d)}\, |a_1+a_2|^2 \nonumber
\end{eqnarray}
where $a_1$, $a_2$ are QCD coefficients \cite{BBNS,BBNS3}.
The advantage of this test is that $B^-\to\pi^-\pi^0$ receives neither
penguin nor annihilation contributions. It thus gives information
on the other aspects of the QCD dynamics in $B\to\pi\pi$.
This test was discussed recently in \cite{LR,BN}.

\item Direct CP asymmetries.
From the heavy-quark limit one generally expects strong phases
to be suppressed, except for a few special cases. 
This circumstance should suppress direct
CP asymmetries. Of course those also depend sensitively on weak phases
and a detailed analysis has to consider individual channels.
At present, qualitatively, one may at least say that
the non-observation of direct CP violation in $B$ decays until today,
with experimental bounds typically at the $10\%$ level,
are not in contradiction with the theoretical expectation.

\item Weak annihilation.
Amplitudes from weak annihilation represent power
suppressed corrections, which are uncalculable in QCD
factorization and so far need to be estimated relying on
models.
At present there are no indications that annihilation terms
would be anomalously large, but they do contribute to the theoretical
uncertainty. Effectively, annihilation corrections may be considered
as part of the penguin amplitudes. To some extent, therefore, they
are tested with the help of the penguin-to-tree ratio discussed above. 
Nevertheless, in order to disentangle
their impact from other effects it is of great interest
to test annihilation separately. This can be done with decay
modes that proceed through annihilation or at least have a dominant
annihilation component.

An example is the pure annihilation channel $B_d\to D^-_s K^+$.
Even though this case is
somewhat different from the reactions of primary interest here,
because of the charmed meson in the final state, it is still useful
to cross-check the typical size of annihilation expected in model
calculations. Treating the $D$ meson in the model estimate for annihilation
\cite{BBNS3} as suggested in \cite{BBNS}, one finds a central
(CP-averaged) branching ratio of  $B(B_d\to D^-_s K^+)=1.2\times 10^{-5}$.
Allowing for a $100\%$ uncertainty of the central annihilation
estimate, which in the case of the penguin-to-tree ratio shown in 
Fig. \ref{fig:pt} corresponds to the inner (solid) error region
around the theoretical value (marked by the cross),
gives an upper limit \cite{BCKMWS} of $5\times 10^{-5}$. This is in agreement
with the current experimental result $(3.8\pm 1.1)\times 10^{-5}$
(see refs. in \cite{BCKMWS}).

Additional tests should come from annihilation decays into
two light mesons, such as $B\to KK$ modes.\cite{GR98}
These, however, are CKM suppressed and only upper limits are known
at present.
The $K^+\bar K^0$ and $K^0\bar K^0$ channels have both
annihilation and penguin contributions. On the other hand
$B\to K^+ K^-$ is a pure weak annihilation process and therefore
especially important.
Further discussions can be found in \cite{BBNS3,BN,GR98}.
\end{enumerate}

At present, within current experimental and theoretical
uncertainties, there are no clear signals of significant
discrepancies between measurements and SM expectations
in hadronic $B$ decays, neither with respect to QCD calculations
nor suggesting the need for new physics. 
However, a few experimental results have central values
deviating from standard predictions, which attracted some
attention in the literature. 
Even though the discrepancies are not significant at the moment, it will
be interesting to follow future developments.
We comment on some of those possible hints here, 
with a view on QCD predictions within the SM.
\begin{itemize}
\item
As seen in (\ref{scwa}) the measurement of $C=-0.38\pm 0.16$ suggests the
possibility of large direct CP violation in $B\to\pi^+\pi^-$
decays. On the other hand, this is largely due to the result from Belle,
whereas BaBar gives a smaller effect.
In the SM one expects $C\approx 0.1$ with an error of about the same size.
It is interesting to note that the perturbative strong interaction phase
predicted to lowest order in QCD factorization gives a positive
value for $C$ while the measurements seem to prefer negative values.
Since the strong phase is a small effect in the heavy-quark limit,
uncalculable power corrections could possibly compete with the
perturbative contribution. A small negative $C$ is therefore not
excluded, but the reliability of a lowest order perturbative
calculation of the strong phase would then be in doubt.
(A logical possibility for $C<0$ would be that the positive sign of the
strong phase is correct, but the weak phase is negative, which would
require new physics in $\varepsilon_K$.)
In any case, a clarification of the experimental situation will
be important. It may also be noted that the central
numbers from Belle, which are large for both $S$ and $C$, would
violate the absolute bound $S^2+C^2 \leq 1$ when taken at face value.

\item
Mixing-induced CP violation $S$ in $B\to\phi K_S$ and $B\to\eta' K_S$,
which proceed through the penguin transition $b\to s\bar ss$,
could be strongly affected by new physics. In the SM one expects
$S_{\phi K_S}$ and $S_{\eta' K_S}$ to be close to the benchmark
observable $S_{\psi K_S}$ of mixing-induced CP violation in
$B\to\psi K_S$.\cite{GIW} Hints of deviations in the data from Belle,
and to a much lesser extent from BaBar, have motivated 
several analyses in the literature on this issue.\cite{GH,DCS,GLNQ}
Experimentally one finds for the world average \cite{TB} 
\begin{eqnarray}
S_{\phi K_S} - S_{\psi K_S} &=& -0.89\pm 0.33 \\
S_{\eta' K_S} - S_{\psi K_S} &=& -0.47\pm 0.22
\end{eqnarray}
where the first result combines the BaBar and Belle values
ignoring the rather poor agreement between them.
This can be compared with the SM expectation based on a recent
QCD analysis in \cite{BN}
\begin{eqnarray}
S_{\phi K_S} - S_{\psi K_S} &=& 0.025\pm 0.016 \\
S_{\eta' K_S} - S_{\psi K_S} &=& 0.011\pm 0.013
\end{eqnarray}
More information on possible new physics implications can be
found in \cite{YG}.

\item
Current data for the ratio of $B\to\pi^+\pi^-$ and $B\to\pi^+\pi^0$
branching fractions appear to be somewhat low in comparison
with theoretical calculations for a CKM phase $\gamma< 90^\circ$
as given by standard fits of the CKM  unitarity triangle.
This feature is often interpreted \cite{HSW} as a hint for a larger
value of $\gamma > 90^\circ$. Such a value could change a
constructive interference of tree and penguin amplitudes in the
$\pi^+\pi^-$ mode into a destructive one, and thus reduce the
ratio of branching fractions. In \cite{BN} a different, QCD related
possibility was discussed that could account for the suppression
of $B\to\pi^+\pi^-$ relative to $B\to\pi^+\pi^0$, even for 
$\gamma< 90^\circ$. In this scenario, which can be realized
without excessive tuning of input parameters, the factorization
coefficient $a_2$ (color-suppressed tree) is enlarged, while 
the $B\to\pi$ form factor is somewhat smaller than commonly assumed.
This keeps $B\to\pi^+\pi^0$ roughly constant and suppresses
$B\to\pi^+\pi^-$, which is independent of $a_2$.
The factorization test mentioned in point 2. above would be very
useful to check such a scenario. This could also help
to clarify the situation with $B\to\pi^0\pi^0$, which is very
sensitive to $a_2$ and for which first measurements from BaBar and
Belle indicate a substantial branching fraction.\cite{JF}
Theoretically $a_2$ is subject to sizable uncertainties, because
color suppression strongly reduces the leading order value 
and makes the prediction sensitive to subleading corrections.

\item
The ratio (CP averaged rates are understood)
\begin{equation}
R_{00}=\frac{2\Gamma(\bar B^0\to\pi^0\bar K^0)}{\Gamma(B^-\to\pi^-\bar K^0)}
\end{equation}
appears to be larger than expected theoretically. This is shown
in Fig. \ref{fig:r00}.
\begin{figure}
\center
\psfig{figure=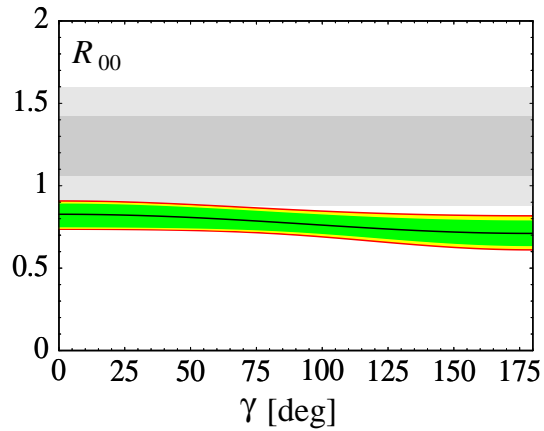, width=8truecm}
\caption[]{Theoretical prediction for 
$R_{00}=2\Gamma(B\to\pi^0 K^0)/\Gamma(B\to\pi^\pm K^0)$.\cite{BN} 
The experimental result is indicated by the straight horizontal
bands showing the $1\sigma$ (dark) and $2\sigma$ (light) range.}
\label{fig:r00}
\end{figure}
The ratio $R_{00}$ is almost insensitive to the CKM angle $\gamma$
and it is essentially impossible to enhance the prediction
in the SM by QCD effects.\cite{BN}
The discrepancy of about $2\sigma$ can also be seen in a
different way, using the Lipkin-Gronau-Rosner sum rule,
which relates all four $\pi K$ modes using isospin symmetry.\cite{LGR}
The ratio
\begin{equation}
R_L=\frac{2\Gamma(\bar B^0\to\pi^0\bar K^0)+2\Gamma(\bar B^-\to\pi^0 K^-)}{
\Gamma(B^-\to\pi^-\bar K^0)+\Gamma(\bar B^0\to\pi^+\bar K^-)}
\end{equation}
can be shown to be 1 up to corrections of se\-cond order in small
quantities. Experimentally it is also about $2\sigma$ high.\cite{GR03}
If the discrepancy should become statistically significant, it
would be a strong indication of physics beyond the SM.\cite{BN,YG,GR03,BFRS}

\end{itemize}

The status of QCD calculations for $B\to PV$ modes
is presented in \cite{BN}
and a more general discussion of new physics aspects is 
given by \cite{YG}.

\section{Rare and Radiative $B$ Decays}

\subsection{Radiative Decays $B\to V\gamma$}

Factorization in the sense of QCD can also be applied to the
exclusive radiative decays $B\to V\gamma$ ($V=K^*$, $\rho$).
The factorization formula for the operators in the
effective weak Hamiltonian can be written as \cite{BFS,BB}
\begin{eqnarray}\label{fform}
&& \langle V\gamma(\epsilon)|Q_i|\bar B\rangle = \\
&& \Bigl[ F^{B\to V}(0)\, T^I_{i} 
+\int^1_0 d\xi\, dv\, T^{II}_i(\xi,v)\, \Phi_B(\xi)\, \Phi_V(v)\Bigr]
\cdot\epsilon \nonumber
\end{eqnarray}
where $\epsilon$ is the photon polarization 4-vector.
Here $F^{B\to V}$ is a $B\to V$ transition form factor,
and $\Phi_B$, $\Phi_V$ are leading twist light-cone distribution amplitudes
(LCDA) of the $B$ meson and the vector meson $V$, respectively.
These quantities 
describe the long-distance dynamics of the matrix elements, which
is factorized from the perturbative, short-distance interactions
expressed in the hard-scattering kernels $T^I_{i}$ and $T^{II}_i$.
The QCD factorization formula (\ref{fform}) holds up to
corrections of relative order $\Lambda_{QCD}/m_b$.
Annihilation topologies are power-suppressed, but still calculable
in some cases. The framework of QCD factorization is necessary to
compute exclusive $B\to V\gamma$ decays systematically beyond the
leading logarithmic approximation. Results to next-to-leading order
in QCD, based
on the heavy quark limit $m_b\gg\Lambda_{QCD}$ have been
computed \cite{BFS,BB}
(see also \cite{AP}).

The method defines a systematic,
model-independent framework for $B\to V\gamma$.
An important conceptual aspect of this analysis is the interpretation
of loop contributions with charm and up quarks, which come from
leading operators in the effective weak Hamiltonian.
These effects are calculable in terms of
perturbative hard-scattering functions and universal meson
light-cone distribution amplitudes. They are ${\cal O}(\alpha_s)$
corrections, but are leading power contributions in the
framework of QCD factorization. This picture is in contrast to the
common notion that considers charm and up-quark loop effects as
generic, uncalculable long-distance contributions.
Non-factorizable long-distance corrections may still exist, but
they are power-suppressed.
The improved theoretical understanding of $B\to V\gamma$ decays
streng\-thens the motivation for still more detailed
experimental investigations, which will contribute
significantly to our knowledge of the flavor sector.

\begin{figure}
\center
\psfig{figure=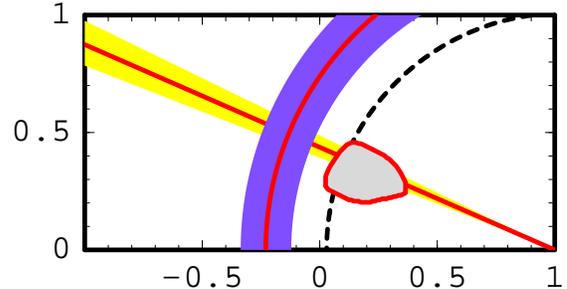, width=8truecm}
\caption[]{Impact of the current experimental upper limit
on $B(B\to\rho\gamma)/B(B\to K^*\gamma)$ in the $(\bar\rho,\bar\eta)$
plane. The area to the left of the dark band is excluded.
The width of the dark band reflects the variation
of $\xi\equiv F_{K^*}/F_\rho=1.33\pm 0.13$ (second ref. in \cite{PBB}).
The case of $\xi=1$ is illustrated by the dashed curve.
The intersection with the light-shaded band from the measurement of
$\sin 2\beta$ defines the apex of the unitarity triangle
and the length of $R_t=\sqrt{(1-\bar\rho)^2 + \bar\eta^2}\sim |V_{td}|$,
once the upper limit will be turned into a measurement.
The irregular area represents the standard unitarity triangle fit.}
\label{fig:rvtd}
\end{figure}

The uncertainty of the branching fractions is 
currently dominated by the form factors $F_{K^*}$, $F_\rho$. 
A NLO analysis \cite{BB} yields 
(in comparison with the experimental results in brackets)
$B(\bar{B}\to \bar{K}^{*0}\gamma)/10^{-5}=7.1\pm 2.5$  
($4.21\pm 0.29$ \cite{BABAR1})
and $B(B^-\to\rho^-\gamma)/10^{-6}=1.6\pm 0.6$ 
($< 2.3$ \cite{BABAR2}).
Taking the sizable uncertainties into 
account, the results for $B\to K^*\gamma$ are compatible with the 
experimental measurements, 
even though the central theoretical values appear to be somewhat high.
$B(B\to\rho\gamma)$ is a sensitive
measure of CKM quantities.\cite{BB,SWB,AL} This is illustrated in
Fig. \ref{fig:rvtd}.

\subsection{SCET}

In decay processes of $B$ mesons with highly energetic light quarks
in the final state, HQET alone is not sufficient to account for the
complete long-distance degrees of freedom that need to be represented
in an effective theory description. A first step towards implementing
the missing ingredients was made in \cite{DG}.
In this paper a framework, called large-energy effective theory (LEET),
was suggested that describes the interactions of energetic light quarks
with soft gluons. To correctly reproduce the infrared structure of QCD,
also collinear gluons need to be included, which was emphasized in
\cite{BFPS}. The authors of \cite{BFPS} constructed an effective
theory, the SCET, for soft and collinear gluons, applicable to
energetic heavy-to-light transitions. These transitions may be
inclusive heavy-to-light processes, such as $b\to u$ decays, but
also exclusive $B\to P$, $V$ form factors at large recoil of the
light final state meson. Similarly the SCET is
a useful language to investigate factorization properties in hadronic
$B$ decays in general terms.

For the construction of the SCET one writes the four-momentum $p$ of 
an energetic light quark (collinear quark) in light-cone coordinates
\begin{equation}\label{ppmpp1}
p^\mu=\frac{1}{\sqrt{2}}\left(p_- n^\mu + p_+ \bar n^\mu\right)+p^\mu_\perp
\end{equation}
\begin{equation}\label{ppmpp2}
p_\pm=\frac{p^0\pm p^3}{\sqrt{2}}
\end{equation}
where $n$ is a light-like four-vector in the direction of the
collinear quark and $\bar n$ is a similar vector in the opposite
direction, that is
\begin{equation}\label{nnbar}
n^2=\bar n^2=0\qquad n\cdot\bar n=2
\end{equation}
The four-vector $p_\perp$ contains the components of $p$ perpendicular
to both $n$ and $\bar n$.
For $p$ collinear to the light-like direction $n$ the components
scale as $p_-\sim M$, $p_\perp\sim M\lambda$, $p_+\sim M\lambda^2$,
where $M$ is the hard scale ($\sim m_b$) and $\lambda$ is a small
parameter, such that $p^2=2p_+ p_-+p^2_\perp\sim M^2\lambda^2$.
The dependence on the larger components of $p$, $p_-$ and $p_\perp$ is then
removed from the light-quark field $\psi(x)$ in full QCD by writing
\begin{equation}\label{ptilde1}
\psi(x)=\sum_{\tilde p}\,  e^{-i\tilde p\cdot x}\, \psi_{n,p}
\end{equation}
\begin{equation}\label{ptilde2}
\tilde p\equiv \frac{1}{\sqrt{2}} p_- n+ p_\perp
\end{equation}
This is analogous to the construction of the HQET, where the dependence
on the large components $v$ of the heavy-quark velocity is isolated
in a similar way. The new fields $\psi_{n,p}$ are then projected
onto the spinors
\begin{equation}\label{xins}
\xi_{n,p}=\frac{\not\! n \not\!\bar n}{4}\, \psi_{n,p}\qquad
\xi_{\bar n,p}=\frac{\not\!\bar n \not\! n}{4}\, \psi_{n,p}
\end{equation}
The field $\xi_{n,p}$ represents the collinear quark in the effective
theory. The smaller components $\xi_{\bar n,p}$ are integrated out
in the construction of the effective theory Lagrangian ${\cal L}_{SCET}$
from the Lagrangian of full QCD. ${\cal L}_{SCET}$ contains collinear
quarks $\xi_{n,p}$, the heavy-quark fields from HQET, $h_v$, and
soft and collinear gluons.

A typical application is the analysis of $B\to P$, $V$ form factors
at large recoil. Bilinear heavy-to-light currents $\bar q\Gamma b$
have to be matched onto operators of the SCET, schematically
\begin{equation}\label{qgammab}
\bar q\Gamma b\to C_i\, \bar\xi_{n,p}\tilde\Gamma_i h_v
\end{equation}
where the $C_i$ are Wilson coefficient functions. For
$B\to P$, $V$ transitions in full QCD there is a total of ten different
form factors describing the matrix elements of the possible
independent bilinear currents. In SCET the equations of motion
\begin{equation}\label{sceteom}
\not\! v h_v= h_v \qquad \not\! n\xi_{n,p}=0
\end{equation}
imply constraints, which reduce the number of independent
form factors to three, to leading order in the heavy-quark limit.
An application to $B\to K^* l^+l^-$ decays will be discussed
in the following section.
Further developments and applications of the SCET framework to
rare, radiative and hadronic $B$ decays can be found in
\cite{BCDF,BF02,BF03,HN,HLPW}.

\subsection{Forward-Backward Asymmetry Zero in $B\to K^* l^+l^-$}

Substantial progress has taken place over the last few years
in understanding the QCD dynamics of exclusive $B$ decays.
The example of the forward-backward asymmetry in $B\to K^* l^+l^-$
nicely illustrates some aspects of these developments.

\begin{figure}
\center
\vspace{3cm}
\hspace{-10cm}
\psfig{figure=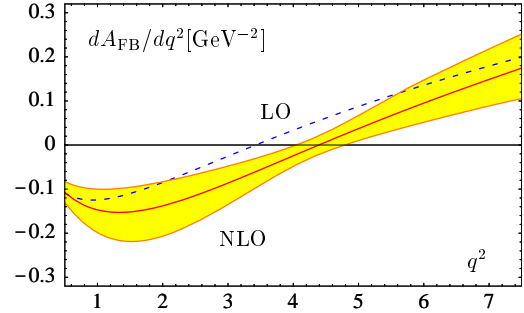, width=3truecm}
\caption[]{$A_{FB}$ spectrum for $\bar B\to K^*l^+l^-$ at
leading and next-to-leading order in QCD.\cite{BFS}}
\label{fig:afb}
\end{figure}

The forward-backward asymmetry $A_{FB}$ is the rate difference
between forward ($0<\theta <\pi/2$) and backward ($\pi/2 <\theta < \pi$)
going $l^+$, normalized by the sum, where $\theta$ is the angle between
the $l^+$ and $B$ momenta in the centre-of-mass frame of the dilepton pair.
$A_{FB}$ is usually considered as a function of the dilepton mass $q^2$.
In the standard model the spectrum $dA_{FB}/dq^2$ (Fig. \ref{fig:afb})
has a characteristic zero at
\begin{equation}\label{afb0}
\frac{q^2_0}{m^2_B}=-\alpha_+ \frac{m_b C_7}{m_B C^{eff}_9}
\end{equation}
depending on short-distance physics contained in the coefficients
$C_7$ and $C^{eff}_9$. The factor $\alpha_+$, on the other hand,
is a hadronic quantity containing ratios of form factors.

It was first stressed in \cite{BUR} that $\alpha_+$ is
not very much affected by hadronic uncertainties and very
similar in different models for form factors with $\alpha_+\approx 2$.
After relations were found between different heavy-light form factors 
($B\to P$, $V$)
in the heavy-quark limit and at large recoil \cite{CLOPR},
it was pointed out in \cite{ABHH} that as a consequence $\alpha_+ =2$ holds  
exactly in this limit. Subsequently, the results of \cite{CLOPR}
were demonstrated to be valid beyond tree level \cite{BFS,BFPS}.
The use of the $A_{FB}$-zero  as a {\it clean\/} test of
standard model flavor physics was thus put on a firm basis and
NLO corrections to (\ref{afb0}) could be computed \cite{BFS}.
More recently also the problem of power corrections to heavy-light form
factors at large recoil in the heavy-quark limit has been studied
\cite{BCDF}.
Besides the value of $q^2_0$, also the sign of the slope
of $dA_{FB}(\bar B)/dq^2$ can be used as a probe of new physics.
For a $\bar B$ meson, this slope is predicted to be positive
in the standard model \cite{BHI}. 

\subsection{Radiative Leptonic Decay $B\to l\nu\gamma$}

\begin{figure}
\center
\psfig{figure=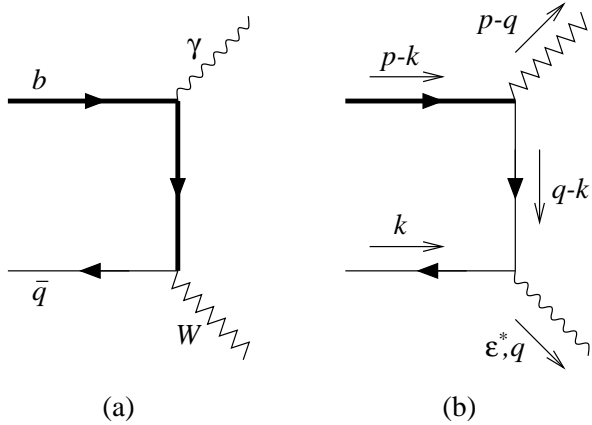, width=8.0truecm}
\caption[]{Tree-level diagrams for $B\to l\nu\gamma$.
Only diagram (b) contributes at leading power.\cite{DGS}}
\label{fig:blnug}
\end{figure}

The tree-level process $B\to l\nu\gamma$ is not so much of
direct interest for flavor physics, but it provides us with an 
important laboratory for studying QCD dynamics in exclusive
$B$ decays that is crucial for many other applications.
The leading-power contribution comes from the diagram
in Fig. \ref{fig:blnug} (b), 
which contains a light-quark
propagator that is off-shell by an amount $(q-k)^2\sim q_- k_+$
Here $q$ is the hard, light-like momentum of the photon
with components scaling as $m_b$ (this restricts the region of phase-space
where the present discussion applies), and $k$ is the soft momentum of
the spectator quark. The decay is thus determined by a hard-scattering
process, but also depends on the structure of the $B$ meson in a
non-trivial way \cite{KPY}. 
Recently, in \cite{DGS} it has been proposed, and shown to one loop in QCD, 
that the form factors $F$ for this decay factorize as
\begin{equation}
F=\int d\tilde k_+ \Phi_B(\tilde k_+)\, T(\tilde k_+)
\end{equation}
where $T$ is the hard-scattering kernel and $\Phi_B$ the light-cone
distribution amplitude of the $B$ meson defined as
\begin{equation}
\Phi_B(\tilde k_+)=\int dz_- e^{i\tilde k_+ z_-}
\langle 0|b(0) \bar u(z)|B\rangle|_{z_+=z_\perp=0}
\end{equation}
The hard process is characterized by a scale $\mu_F\sim \sqrt{m_b\Lambda}$.
At lowest order the form factors are proportional to
$\int d\tilde k_+\, \Phi_B(\tilde k_+)/\tilde k_+\equiv 1/\lambda_B$,
a parameter that enters hard-spectator processes in many other
applications. The analysis at NLO requires resummation of
large logarithms $\ln(m_b/\tilde k_+)$.  
An extension of the proof of factorization to all orders was subsequently
given by \cite{LPW,BHLN} within the SCET.

Progress has also been made recently towards a better understanding
of the $B$ meson light-cone distribution 
amplitude itself.\cite{BFS,GN,KKQT,LN,BIK}

\section{Conclusions}

QCD has been very successful as a theory of the strong interaction
at high energies, based on expansions in inverse powers of
the high-energy scale and perturbation theory in $\alpha_s$.
This general framework of QCD has recently found new applications
in the treatment of exclusive decays of heavy hadrons.
It is particularly exciting that these developments come
at a time where a large amount of precision data is being
collected at the experimental $B$ physics facilities.

Factorization formulas in the heavy-quark limit have been
proposed for a large variety of exclusive $B$ decays.
They justify in many cases the phenomenological factorization
ansatz that has been employed in many applications.
In addition they enable consistent and systematic calculations
of corrections in powers of $\alpha_s$. Non-factorizable
long-distance effects are not calculable in general but they are
suppressed by powers of $\Lambda_{QCD}/m_b$. So far,
$B\to D^+\pi^-$ decays are probably understood best. Decays
with only light hadrons in the final state such as $B\to\pi\pi$,
$K^*\gamma$, $\rho\gamma$, or $K^*l^+l^-$ include hard spectator
interactions at leading power and are therefore more complicated.
An important new tool that has been developed is the soft-collinear
effective theory (SCET), which is of use for proofs of factorization
and for the theory of heavy-to-light form factors at large recoil.
Studies of the process $B\to l\nu\gamma$ have
also led to a better understanding of QCD dynamics in exclusive
hadronic $B$ decays.
These are promising steps towards controling the QCD dynamics in 
exclusive hadronic or rare $B$ decays in a reliable way.
In many cases the required theoretical accuracy is not extremely
high and even moderately precise, but robust predictions
will be very helpful. 
Using all the available tools we can hope to successfully
probe CP violation, weak interaction parameters and new phenomena in the 
quark-flavor sector.

\clearpage
\section*{DISCUSSION}

\begin{description}

\item[Brendan Casey] (Brown University):
Does the range in predictions for $\bar B^0\to D^+_s K^-$ of 
$(1\div 5)\times 10^{-5}$ correspond to the $1\sigma$ contours or to the 
$5\sigma$ contours in the $P/T$ predictions?

\item[Gerhard Buchalla{\rm :}]
The default model estimate for the annihilation term
gives $1.2\times 10^{-5}$ for the branching ratio of $\bar B^0\to D^+_s K^-$.
Allowing for a $100\%$ uncertainty of the default value gives the
upper limit of $5\times 10^{-5}$. This corresponds to the
inner (solid line) of the three error contours shown in the
plot of the $P/T$ prediction (see Fig. \ref{fig:pt}).

\item[Harry Lipkin] (Weizmann Institute):
Do you have anything to say about the $B$ decays to the new charmed-strange 
axial and scalar mesons that have been observed? When I predicted last year 
a large $B$ decay to the $D^*_s$ axial vector, I was told by HQET experts 
that this decay would be small.

\item[Gerhard Buchalla{\rm :}]
The $D^*_s$ emitted in $B$ decay is a heavy-light meson and therefore 
represents an extended hadronic object, in contrast to a pion or
a similar energetic light meson. The usual factorization formulas
do not apply to this situation and it is thus difficult to
control QCD uncertainties in the predictions. 

\item[Ikaros Bigi] (Notre Dame University):
When you consider $B\to VV$, like $B\to\rho\rho$, and calculate the 
polarization of $V$, there are corrections of order $1/m_b$. 
Those are sensitive to long-distance dynamics, right?

\item[Gerhard Buchalla{\rm :}]
That is correct.

\end{description}

\end{document}